\documentclass[preprint,preprintnumbers,amsmath,amssymb,nofootinbib]{revtex4}
\usepackage{graphicx}
\usepackage{dcolumn}
\usepackage{bm}
\usepackage{color}

\begin{document}
\title{Non-linear Hamiltonian models for DNA}

\author{Marco Zoli}

\affiliation{School of Science and Technology \\  University of Camerino, I-62032 Camerino, Italy \\ marco.zoli@unicam.it}

\date{\today}

\begin{abstract}
Nucleic acids physical properties have been investigated  by theoretical methods based both on fully atomistic representations and on coarse grained models, e.g. the worm-like-chain, taken from polymer physics. In this article, I present an intermediate (mesoscopic) approach and show how to build a three dimensional Hamiltonian model which accounts for the main interactions responsible for the stability of the helical molecules. While the 3D mesoscopic model yields a sufficiently detailed description of the helix at the level of the base pair, it also allows one to predict the thermodynamical and structural properties of molecules in solution. Relying on the idea that the base pair fluctuations can be conceived as trajectories, I have built a computational method based on the time dependent path integral formalism to derive the partition function. While the main features of the method are presented, I focus here in particular on a newly developed statistical method to set the maximum amplitude of the base pair fluctuations, a key parameter of the theory. Some applications to the calculation of DNA flexibility properties are discussed together with the available experimental data.
\end{abstract}

\maketitle

\tableofcontents
\newpage

\section*{1. Introduction}

In living organisms DNA molecules store and carry the genetic information encoded in the sequence specificity of the two complementary strands assembled via the Watson-Crick base pairing. While the specific bonding of base pairs (bps) lies at the heart of the DNA biological functions, 
predictability and thermodynamic properties of the base pairing are also crucial to a variety of DNA-based methodologies such as detection of genomic variations via DNA microarrays \citep{bots08}, classification of genetic distance between species  \cite{wilson75}, mutation scanning and genotyping of Polymerase-Chain Reaction (PCR) products \cite{zhou07} using high-resolution melting  whereby the accuracy of the process clearly relies on the control of the base pairing thermal stability.

The thermal separation of the helical strands has been a focus of research both in molecular biology and biochemistry since the fundamental biological processes of replication, transcription and protein binding require the local unzipping of the double helix  which allows for reading and
copying of the genetic code \cite{croq00}.

While the DNA structure is stable at room temperature mostly due to the covalent bonds between adjacent nucleotides along the sugar-phosphate backbone,  thermal fluctuations can locally disrupt the hydrogen bonds, starting in regions rich in the weaker adenine-thymine
bps, thus leading to  formation of transient breathing bubbles  as the energy scale for the separation of the bonds between paired bases is of a few $ \,k_BT$ , $k_B$ being the Boltzmann constant and $T$ is the temperature. 
While denaturation bubbles generally appear both in linear and circular supercoiled DNA as a response to release the torsional stress, their size and number varies with the ambient conditions, sequence heterogeneity and  chain length.

The denaturation transition of DNA in solution is made evident by the large increase in the UV absorption spectra due to the rearrangement of the $\pi$ electrons of the bases once the stacking of planar adjacent bases decreases and the hydrogen bonds break. The percentage increase in light absorption at $\sim 260$ {nm} is proportional to the relative presence, in the heterogeneous  molecule, of adenine-thymine (AT) \emph{bps} with respect to guanine-cytosine (GC) \emph{bps} 
\cite{wart85}. In fact the latter present three hydrogen bonds while the former, with two hydrogen bonds, can be more easily disrupted by thermal effects. Note that the effective bond energies of AT-\emph{bps} may be $\sim 30$ meV, that is just above $k_BT$ at room temperature.
Thus, the melting temperature, usually defined by the mid-point transition at which half of the bps are broken, provides a measure of the relative content of $GC$ and $AT$ pairs in the sequence \cite{lucia04}. 

Providing an example of a phase transition, the DNA melting has been widely studied in statistical physics \cite{pol66} and a vast literature has been produced to analyze the helix-coil transition together with the thermally driven formation of denaturation bubbles. Also the flexibility properties of DNA helices have been extensively investigated: among them, the force-extension behavior, the persistence length and the cyclization probability i.e., the probability that, for a given ensemble of linear chains, a fraction of them will close into a loop. 
To these purposes, several computational methods such as transfer integrals  \cite{singh11,hando12} and transfer matrix \cite{menon} methods, Monte Carlo simulations \cite{kalos11}, molecular dynamics  \cite{maiti15},  and path integrals \cite{io13,io14a,io16}  have been applied both to mesoscopic Hamiltonian and polymer physics models. The need for models at intermediate length scales arises from the fact that, for instance, the thymine nucleobase contains fifteen atoms and the DNA monomer unit  (the nucleotide made of a nucleobase plus sugar-phosphate group) contains a few tens of atoms \cite{calla}. Hence, even for a short double stranded DNA sequence, a fully atomistic representation becomes computationally extremely time consuming and unsuitable to derive quantitative information on specific physical properties.

Among the mesoscopic Hamiltonian approaches,  the Peyrard-Bishop (PB) model \cite{pey89} has been widely used in DNA investigations over the last decades. 
The PB Hamiltonian assumes a point-like representation for the nucleotides and describes the main forces which stabilize the DNA molecule, hydrogen bonds between inter-strand pair mates and intra-strand harmonic stacking between adjacent bases, 
in terms of a single degree of freedom, that is the relative distance between pair mates. 

It follows that the model is essentially one-dimensional and, under specific conditions to be stated below, it maps onto an exactly solvable Schr\"{o}dinger equation which yields a crossover temperature characteristic of a smooth thermal denaturation. A later version of the PB model   (termed the PBD model \cite{pey93}) incorporating an-harmonic stacking interactions has been proposed to account for the sharpness of the melting transition (in the thermodynamic limit of an infinite chain) although the character of the denaturation generally depends on the length and specificities of the sequence \cite{io10}. 
Furthermore, the PB and PBD model treat the base pair (bp) displacements as continuous variables  thus accounting for the dynamics of the nucleotides on complementary strands through an effective Morse potential for the hydrogen bonds between the bp mates. 
While this property allows in principle for a description of those intermediate states which are relevant to the DNA dynamics, the PBD model predictions for the bp lifetimes of open and closed states have yielded much shorter estimates than those inferred from proton–deuterium exchange experiments \cite{gueron85}. 
Accordingly,  improved versions of the 1D model \cite{pey09}  have added solvent potential terms which \textit{a)} enhance the dissociation energy over the Morse plateau and \textit{b)} introduce a hump whose maximum determines the threshold around which a bp may first temporarily open and then either
re-close or fully dissociate. Such re-closing barrier accounts for the hydrogen bonds that the open bases may establish with the solvent, albeit at an higher cost as bases are hydrophobic. Although such improvements partly reconcile the calculated bp lifetimes with the experimental estimates, it is pointed out that the PB model assumes a ladder configuration for the DNA chain whereas a realistic analysis of the thermodynamics and dynamical properties of DNA  should not overlook the helical structure of the molecule with its twisting and bending degrees of freedom \cite{io16b}. Notwithstanding these caveats,  the PB Hamiltonian
still provides a simple and appealing representation for the double stranded chain (which lies at the base of its popularity) and therefore I take it as the initial point of the discussion. 

Beginning with an analysis of the fundamental  properties of the one-dimensional harmonic model, I will follow a step by step procedure to add more physical information to the Hamiltonian model eventually attaining a realistic three-dimensional mesoscopic description for a helical molecule in solution.

Section 2 presents both the harmonic and the an-harmonic one-dimensional Hamiltonian pointing out both the merits and the shortcomings of a ladder model. 
The  three-dimensional Hamiltonian with radial, twisting and bending degrees of freedom is presented in Section 3 together with the effects of a solvent potential. Section 4 outlines the theoretical background for the computational method which I have built to get the structural and thermodynamical properties of DNA molecules in solution. A newly developed statistical approach to set the maximum amplitude for the bps fluctuation in helical molecules is presented and its merits are highlighted.
Section 5 shows an application of the theory to the calculation of the cyclization probability of short DNA chains together with a comparison to available experimental data. Some final remarks are made in Section 6.

\section*{2. 1D Hamiltonian Model}

\renewcommand{\theequation}{2.\arabic{equation}}
\setcounter{equation}{0}

\begin{figure}
\includegraphics[height=8.0cm,width=8.0cm,angle=-90]{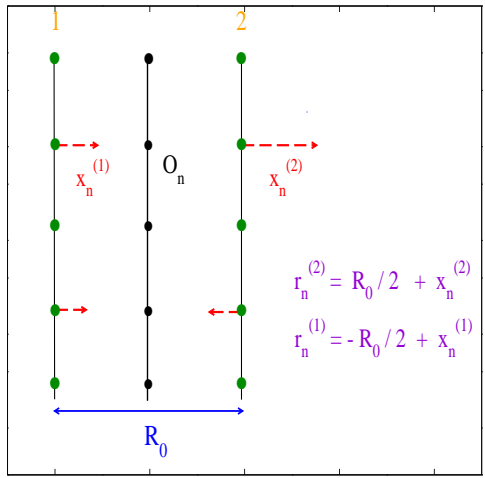}
\caption{\label{fig:1}(Color online)  
Ladder model for an open end chain with $N$ point-like base pairs arranged along the two complementary strands.  The transverse base fluctuations, $r_n^{(1,2)}$, are measured with respect to the mid-chain axis hence, the distance between the pair mates is defined with respect to $R_0$. 
}
\end{figure}

In a fundamental work Englander et al. \cite{engl} suggested that the transient opening of adjacent bps, associated with torsional oscillations around the helix axis, could generate coherent thermally activated soliton excitations propagating along the DNA backbone. These observations have fostered a line of research on the non-linear dynamics of DNA which continues nowadays \cite{yakus}.
Following the observations by Prohofsky \cite{proh} on the strong non-linearities in the hydrogen bonds stretching modes, Peyrard and Bishop \cite{pey89} put forward a minimal harmonic model  to calculate the average inter-strand separation as a function of temperature.
The schematic of the model is shown in Fig.~\ref{fig:1}. The double helix is represented by a ladder
of $N$ point-like homogeneous bps (of common mass  $\mu$)  which are arranged like beads along two parallel strands set at the distance $R_0$. The latter is the bare helix diameter in the absence of fluctuations which, in fact, does not appear explicitly in the original PB model (nor in the PBD). For each pair mate, only transverse fluctuations are considered as these are generally much larger than the longitudinal bps displacements (along the molecule backbone) which are accordingly dropped.

\subsection*{ 2.1 \, Harmonic Stacking Potential}

The Hamiltonian of the model first proposed by Peyrard and Bishop \cite{pey89} reads:

\begin{eqnarray}
& & H^{PB} =\, \sum_{n=1}^N \Biggl[ \frac{\mu}{2} \Bigl( \bigl( \dot{r}_n^{(1)} \bigr)^{2} + \bigl( \dot{r}_n^{(2)} \bigr)^{2} \Bigr) +  V_S\bigl(r_n^{(1)}, r_{n-1}^{(1)}\bigr) +  V_S\bigl(r_n^{(2)}, r_{n-1}^{(2)}\bigr)  +  V_M\bigl(r_n^{(1)} - r_n^{(2)}\bigr)  \Biggr] \, \, , \nonumber
\\
& & V_S\bigl(r_n^{(1)}, r_{n-1}^{(1)}\bigr)=\, \frac{K}{2} \bigl( r_n^{(1)} - r_{n-1}^{(1)}  \bigr)^2 \, \, , \nonumber
\\
& & V_M\bigl(r_n^{(1)} - r_n^{(2)}\bigr)=\, D \biggl[\exp\bigl( -a (r_n^{(1)} - r_n^{(2)} ) \bigr) - 1 \biggr]^2  \, .
\label{eq:0}
\end{eqnarray}

The model contains: \textit{i)} a two particle harmonic intra-strand stacking potential with force constant $K$ and \textit{ii)} a one particle inter-strand Morse potential which represents the hydrogen bonds between pair mates. $D$ is the bp dissociation energy and $a$ is the inverse length setting the range of the Morse potential. While the latter is a usual choice to model hydrogen bonds, any other potential with a
hard core accounting for the  electrostatic repulsion between negatively charged phosphate groups on complementary strands, a stable minimum and a dissociation plateau would be physically suitable. 

Eq.~(\ref{eq:0}) is conveniently rewritten in terms of the variables, \, $z_n=\,(r_n^{(1)} + r_n^{(2)} )/\sqrt{2}$ and $y_n=\,(r_n^{(1)} - r_n^{(2)} )/\sqrt{2}$,
 which describe respectively the in-phase and out-of-phase displacements depicted in Fig.~\ref{fig:1}. Then, the transformed Hamiltonian reads:

\begin{eqnarray}
& & H^{PB} =\, \sum_{n=1}^N \Biggl[ \frac{q_{n}^2}{2\mu} + \frac{K}{2} \Bigl( z_n - z_{n-1}  \Bigr)^2  +  \frac{p_{n}^2}{2\mu} + \frac{K}{2} \Bigl( y_n - y_{n-1}  \Bigr)^2  + D \Bigl[\exp\bigl( -\bar{a} y_n \bigr) - 1 \Bigr]^2 \Biggr] \, \, , \nonumber
\\
& & q_n \equiv \mu \dot{z}_n \, \, , \nonumber
\\
& & p_n \equiv \mu \dot{y}_n  \, \, , \nonumber
\\
& & \bar{a} \equiv a\sqrt{2} \, .
\label{eq:0a}
\end{eqnarray}

\begin{figure}
\includegraphics[height=8.0cm,width=8.0cm,angle=-90]{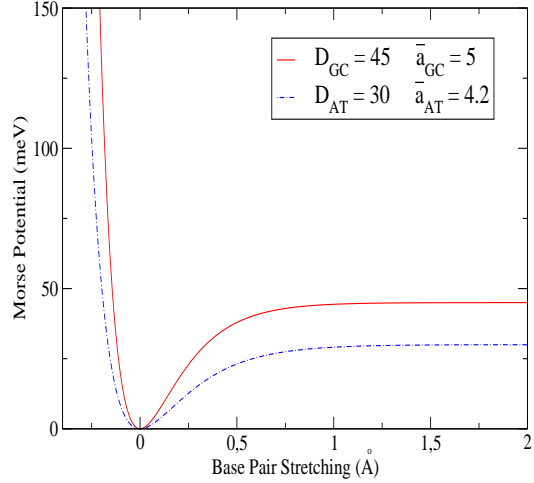}
\caption{\label{fig:2}(Color online)  Morse potential $V_M$ versus base pair separation, calculated from Eq.~(\ref{eq:0}) after setting $R_0=\,0$. Two sets of parameters, suitable for GC and AT base pairs are considered. $D_n$ is in units meV and $\bar{a}_n$ is in \AA$^{-1}$.
}
\end{figure}

While Eq.~(\ref{eq:0a}) is quadratic in the in-phase coordinate, the non-linear contributions are ascribed to the out-of-phase coordinate which stretches the hydrogen bonds. 
As an example, Fig.~\ref{fig:2} shows the Morse potential as a function of the pair mates separation for two parameter choices usually taken to represent GC and AT bps.

The model can be also extended to deal with a chain of heterogeneous bps by introducing site dependent parameters i.e.,  $K \rightarrow K_{n, n-1}$, $D \rightarrow D_n$ and  $\bar{a} \rightarrow \bar{a}_n$.
The thermodynamics of the Hamiltonian in Eq.~(\ref{eq:0a}) can be derived by calculating the classical partition function which, in the canonical ensemble, is given by:

\begin{eqnarray}
& & Z^{PB} =\, \frac{1}{h^{2N}}\prod_{n=1}^N \int_{-\infty }^{\infty } dq_{n} \int_{-\infty }^{\infty } dz_{n}  \int_{-\infty }^{\infty } dp_{n} \int_{-\infty }^{\infty } dy_{n} \exp\bigl[-\beta  H^{PB}(q_{n}, z_{n}, p_{n}, y_{n}) \bigr] \, , \nonumber 
\\
\label{eq:0b}
\end{eqnarray}

where $\beta =\, 1/(k_BT)$. As a basic notion of statistical mechanics, the partition function must be dimensionless. Accordingly the constant $h$ is introduced to fulfill this requirement. Observing that, for each base, the double integral \, $\int dq_{n} \int dz_{n} $ \, (or $\int dp_{n} \int dy_{n} $ ) has the dimension of energy times time (action), $h$ is consistently taken as the Planck constant. While this choice can be rigorously motivated in quantum mechanics, it may sound somewhat arbitrary in a classical context. Note however that in the calculation of average physical quantities, such as the energy, the pre-factors involving $h$ cancel out.

Inserting Eq.~(\ref{eq:0a}) in Eq.~(\ref{eq:0b}), one can promptly integrate out the momenta and the in-phase contributions so that $Z^{PB}$ transforms into:

\begin{eqnarray}
& & Z^{PB} =\, \biggl(\frac{2\pi \mu }{h^{2} \beta }\biggr)^{N} \biggl(\frac{2\pi }{\beta K}\biggr)^{N/2} \prod_{n=1}^N   \int_{-\infty }^{\infty } dy_{n} \exp\bigl[-\beta ( V_S(y_n, y_{n-1}) + V_{M}( y_{n}) ) \bigr] \, \, , \nonumber
\\
& & V_S(y_n, y_{n-1})=\, \frac{K}{2} ( y_n -  y_{n-1} )^2 \, \, , \nonumber
\\
& & V_M(y_n) =\, D \bigl(\exp(- \bar{a} y_n) - 1 \bigr)^2  \, .
\label{eq:0c}
\end{eqnarray}

Thus the model is essentially one dimensional as it is written in terms of the stretching mode $y_n$  which measures
the distance between pair mates  for the \emph{n-th} bp. While at this stage such distance is generally assumed to vary from $-\infty$ to $\infty$, a consistent method to define a finite cutoff on the base pair fluctuations is given in Section 4.

Also note that $y_n$ should be generally defined with respect to the bare helix diameter but, in the PB model, the absence of fluctuations corresponds to \,  $y_n=\,0$ as in Fig.~\ref{fig:2}.  Further, it is pointed out that the PB model sets to zero also the bare rise distance between adjacent bps along the chain.

The multiple integrals in Eq.~(\ref{eq:0c}) can be carried out exactly in the limit of a large system, $N \rightarrow \infty$, assuming periodic boundary conditions and making use of Transfer Integral techniques already applied in the study of the thermodynamics of an anharmonic system with quartic potential \cite{scala}.
This permits to derive the thermodynamical properties once the eigenvalues and eigenvectors of the Transfer Integral operator are known \cite{zhang97}.

Moreover, the PB model can also be studied analytically in the continuum approximation. The latter holds if the intra-strand stacking forces provide the major contribution to the chain stability while the transverse fluctuations are small. In this limit the effective force constant of the Morse potential is  $D \bar{a}^2$  hence, the strong coupling regime is defined by \, $D \bar{a}^2 \ll K$. Under these conditions, a description of the thermally driven separation of the DNA complementary strands, albeit qualitative, can be obtained.
More precisely, the  statistical mechanics of the classical model in Eq.~(\ref{eq:0}) can be mapped onto the quantum mechanics of the Morse oscillator \cite{io14c}.
Solving the Schr\"{o}dinger equation for a particle with mass $\mu$ in the Morse potential \cite{landauQM}, one finds the conditions for the existence of a discrete spectrum of localized states and of a continuous spectrum whereby the transition between the two subsets is physically related to the depth of the Morse potential. 
 Thus, the disappearance of the last bound state in  the Schr\"{o}dinger equation is formally equivalent to the melting of the double helix whose transition temperature is found to be:
 
\begin{eqnarray}
T_c =\, \frac{2}{k_B \bar{a}} {\sqrt{2 K D}} \, \, .
\label{eq:25}
\end{eqnarray}

This picture provides an appealing description of the thermally induced separation between the strands in Fig.~\ref{fig:1} but it should be checked whether the derived expression for $T_c$ and the underlying assumptions i.e., strong coupling and continuum regime, are consistent with the available experimental information. Although large discrepancies are found in the literature regarding the estimate of the harmonic elastic constant in DNA sequences \cite{eijck11}, $K$ values in the range $[20 - 60] meV \cdot$ \AA$^{-2}$ are considered appropriate to most cases and accordingly assumed in mesoscopic models. While one may indeed adjust the parameters in Eq.~(\ref{eq:25}) so as to get $T_c$ of order $\sim  [320 - 370]$ K as estimated for DNA of variable sequence and length \cite{owcz04,wart85}, the fact remains that the above mentioned strong coupling (continuum) hypothesis, \, $D \bar{a}^2 \ll K$, is never verified for meaningful choices of the parameters (see e.g. Fig.~\ref{fig:2}). This suggests that the effects related to the discreteness of the chain are relevant in DNA. For these reasons, theoreticians recur to mesoscopic models in order to account for a description of the molecules at the level of the bp. 

Before proceeding to develop the mesoscopic Hamiltonian, I emphasize that the presence of the on-site Morse potential in Eqs.~(\ref{eq:0c}) (or of any other similar hard-core potential with a plateau) has two fundamental consequences:

1) For very large bps separation, \, $V_S(y_n, y_{n-1})$ grows as $y_n^2$ and the integrand in the first of Eqs.~(\ref{eq:0c}) tends to zero. This however does not occur if all $y_n$ are equal to each other:  in this case,  it is $V_S$ that vanishes whereas the integrand remains finite (that is, it approaches a constant value). Accordingly,  the partition function diverges for \, $y_n \rightarrow \infty$.
The divergence of the partition function arising from the translational mode is well known in physics and also the technique to tackle the divergence is  known \cite{schulman}.
For instance, in semiclassical methods for translationally invariant $\phi^4$ potential models, the zero mode eigenvalue corresponds to the ground state of the fluctuation spectrum. As such mode breaks the Gaussian approximation, it can be extracted from the determinant of the quantum fluctuation around the classical path  and regularized \cite{io07}. Here however this technique does not work, precisely because the on-site $V_M(y_n)$ breaks the translational invariance of the system. As a consequence, the  partition function still diverges for large and equal $y_n$'s while  no straightforward analytic method is available to remove such divergence. For these reasons, a truncation of the phase space available to bps fluctuations is always required in the computational methods applied to Eq.~(\ref{eq:0c}). On the other hand, the cutoff on the bps separations is not only a numerical requirement but it also has physical motivations as there is no reason to assume that the distance between pair mates may get infinitely large for DNA molecules in solution. This issue will be analyzed in detail in Section 4.

2) A theorem due to van Hove \cite{hove} shows that, for 1D models with short range pair interactions such that the partition function is expressed in terms of difference of particle coordinates, the free energy does not have any singularities hence, phase transitions are forbidden in these models.
However, Eq.~(\ref{eq:0c}) contains the unbound on site potential $V_M(y_n)$ acting as an external field, therefore the van Hove's theorem does not apply.

It is worth noticing that neither the general argument given by Landau \cite{landau}, regarding the impossibility of phase transitions in 1D, applies to our system. In fact, this argument states that phase coexistence cannot occur in 1D at finite temperatures as the energetic cost of making a domain wall between two regions is finite.  Instead, in the continuum limit, Eq.~(\ref{eq:0c}) admits a domain wall solution connecting open and closed parts of the molecule but the domain wall energy is infinite for a large system.

Thus, the thermally driven separation of the complementary strands provides an example of phase transition, at least in the thermodynamic limit of an infinite chain, whose properties have been the subject of intense debate \cite{cule97,santos10}.

\subsection*{ 2.2 \, Anharmonic Stacking Potential}

A significant improvement over the PB model  has been brought about by the inclusion of a non-linear term in the stacking potential 
to account for the observed sharpness of the melting process. This is ascribed to the cooperative character of the bp opening along the stack \cite{pey93}. Specifically, in the PBD model, the two particles harmonic potential in Eq.~(\ref{eq:0c}) is replaced by the anharmonic potential

\begin{eqnarray}
& &V_S^{an}(y_n, y_{n-1})=\, \frac{K}{2} \Bigl[ 1 + \rho \exp\bigl[-\alpha(y_n + y_{n-1})\bigr] \Bigr] ( y_n - y_{n-1} )^2 \, \, ,
\label{eq:26}
\end{eqnarray}

where the non-linear parameters $\rho$ and  $\alpha$ drive the cooperative behavior of the system towards denaturation. At this stage the choice of these new parameters may look somewhat arbitrary as they cannot be straightforwardly related to physical observables. However, as for $\alpha$,  the condition \, $\alpha  < \bar{a}$ should be fulfilled in order to ensure that the range of the stacking is taken larger than that of the Morse potential.
This is consistent with the fact that covalent bonds along the stack are stronger
than bp hydrogen bonds hence, a large amplitude transverse fluctuation sampling the Morse plateau may not suffice to unstack the bp. 
In general, when all bps are closed, $y_n \,, \,y_{n-1} \ll \alpha^{-1}$ for all $n$. Under these conditions, we see from Eq.~(\ref{eq:26}) that the effective coupling is $ \sim K(1 + \rho)$. 

If, however, a fluctuation causes either $y_n > \alpha^{-1}$ or $y_{n-1} > \alpha^{-1}$, then the
hydrogen bond between pair mates loosens and the base moves out
of the strand axis. Accordingly the $\pi$ electrons overlap in the base plateaus is reduced, the binding between neighboring bases along the
strand weakens and the effective coupling drops to $\sim K$. As a consequence, also the adjacent base moves out of the stack thus propagating the fluctuational opening.
This explains the link between anharmonicity and cooperativity leading to bubble formation and eventually to denaturation in the anharmonic PBD model.

Then, small $\alpha$ values  indicate that large fluctuations are required to unstack a bp and produce a consistent reduction in the stacking energy while $\rho$ weighs the energetic reduction in going from a closed, stiff bp conformation to the open one. 

Despite the new features introduced by the non-linear terms, it is remarked that:\textit{ i)} neither 
$V_S^{an}(y_n, y_{n-1})$ accounts for the fact that the stacking energy should remain finite also for very large \, $y_n - y_{n-1}$ \, i.e., when adjacent bps along the stack slide past each other until they no longer overlap. In fact, for very large bps separation, $V_S^{an}(y_n, y_{n-1})$ behaves like $V_S(y_n, y_{n-1})$ ; \textit{ii)}  the above discussed problem concerning the divergence of the partition function persists also with the PBD model \cite{io12}.

\section*{3. 3D Hamiltonian Model with Solvent Potential}

\renewcommand{\theequation}{3.\arabic{equation}}
\setcounter{equation}{0}

Following these considerations, it is clear that a more realistic representation of the helical molecule has to go beyond the 1D ladder model and incorporate  the essential degrees of freedom for the bps in a linear chain. In Fig.~\ref{fig:3}, I report  the schematic for the three-dimensional mesoscopic model which has been first presented in ref.\cite{io16b} and there used to calculate the probability for an open ends chain to close into a loop. Essentially, the model incorporates twisting and bending fluctuations for any dimer in the chain assuming that the twist angle $\theta_n$ and the bending angle $\phi_n$ are site dependent. Furthermore, the radial fluctuations $r_{n}$ are defined with respect to the average helix diameter while the finite distance $d$ along the helical stack (neglected in Section 2) now shows up in the model. Straightforward geometrical considerations lead to write the distance between adjacent bps represented by the blue dots in Fig.~\ref{fig:3},  e.g. the segment $\overline{AB}$,  as:

\begin{figure}
\includegraphics[height=8.0cm,width=8.0cm,angle=-90]{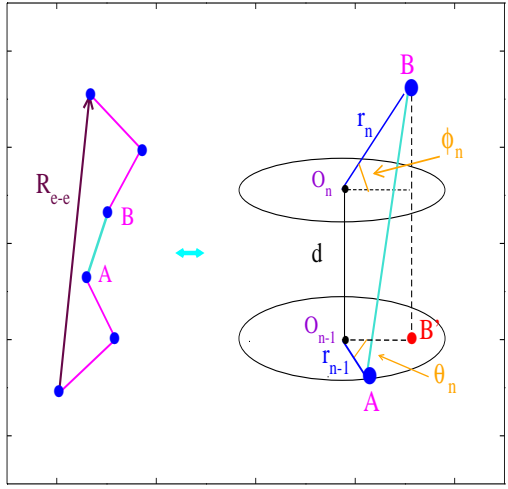}
\caption{\label{fig:3}(Color online)  Three dimensional model for an open end chain with $N$ point-like base pairs.  The segment $\overline{AB}$, i.e. the separation between two neighboring base pairs along the molecule stack, is the distance between the tips of the radial displacements $r_{n}$, $r_{n-1}$. The $r_{n}$'s represent the inter-strand fluctuational distance between the two mates of the $n-th$ base pair. Such distance is measured with respect to the $O_n$'s which lye along the central axis of the helix. $\theta_n$ and $\phi_{n}$ are the twisting and bending angles respectively formed by adjacent base pair fluctuations. 
In the absence of radial fluctuations, all $r_{n}$'s would be equal to the bare helix diameter and the model would reduce to a freely jointed chain model made of $N-1$ bonds, all having length $d$.    In the absence of bending fluctuations, the model would reduce to a fixed-plane representation as depicted by the ovals in the r.h.s. drawing. The latter also convey the idea that the $r_{n}$'s represent in-plane fluctuations. The global size of the molecule is measured by the end-to-end distance, $R_{e-e}$, shown on the l.h.s. drawing.
}
\end{figure}

\begin{eqnarray}
& & \overline{d_{n,n-1}}=\, \bigl[ (d + r_n \sin \phi_n)^{2} + r_{n-1}^2 + (r_n \cos \phi_n)^2 -2 r_{n-1} \cdot r_n \cos \phi_n \cos \theta_n \bigr]^{1/2}  \, \, . 
\label{eq:28}
\end{eqnarray}

This distance appears in the stacking potential with first neighbors interactions of the 3D Hamiltonian for a linear chain whose full expression reads:

\begin{eqnarray}
& &H =\, H_a[r_1] + \sum_{n=2}^{N} H_b[r_n, r_{n-1}, \phi_n, \theta_n] \, , \nonumber
\\
& &H_a[r_1] =\, \frac{\mu}{2} \dot{r}_1^2 + V_{M}[r_1] + V_{sol}[r_1] \, , \nonumber
\\
& &H_b[r_n, r_{n-1}, \phi_n, \theta_n]= \,  \frac{\mu}{2} \dot{r}_n^2 + V_{M}[r_n] +  V_{sol}[r_n] + V_{S}[ r_n, r_{n-1}, \phi_n, \theta_n]  \, \, , \nonumber
\\ 
& &V_{M}[r_n]=\, D_n \bigl[\exp(-\bar{a}_n (|r_n| - R_0)) - 1 \bigr]^2  \, , \nonumber
\\
& &V_{Sol}[r_n]=\, - D_n f_s \bigl(\tanh((|r_n| - R_0)/ l_s) - 1 \bigr) \, , \nonumber
\\ 
& &V_{S}[ r_n, r_{n-1}, \phi_n, \theta_n]=\, K_{n, n-1} \cdot \bigl(1 + G_{n, n-1}\bigr) \cdot \overline{d_{n,n-1}}^2   \, , \nonumber
\\
& &G_{n, n-1}= \, \rho_{n, n-1}\exp\bigl[-\alpha_{n, n-1}(|r_n| + |r_{n-1}| - 2R_0)\bigr]  \, . 
\label{eq:29}
\end{eqnarray}

Thus $V_S$ includes the angular variables through the squared distance in  Eq.~(\ref{eq:28}) and represents the extension to the 3D model of the two particles stacking potential defined in Eq.~(\ref{eq:26}).  

As Eq.~(\ref{eq:29}) will be used in the following calculations, some further comments are deemed necessary:

\textit{(i)}  $H_a[r_1]$ is taken out of the sum as the first bp has only one neighbor  i.e., it is coupled only to the successive bp along the chain.

\begin{figure}
\includegraphics[height=10.0cm,width=10.0cm,angle=-90]{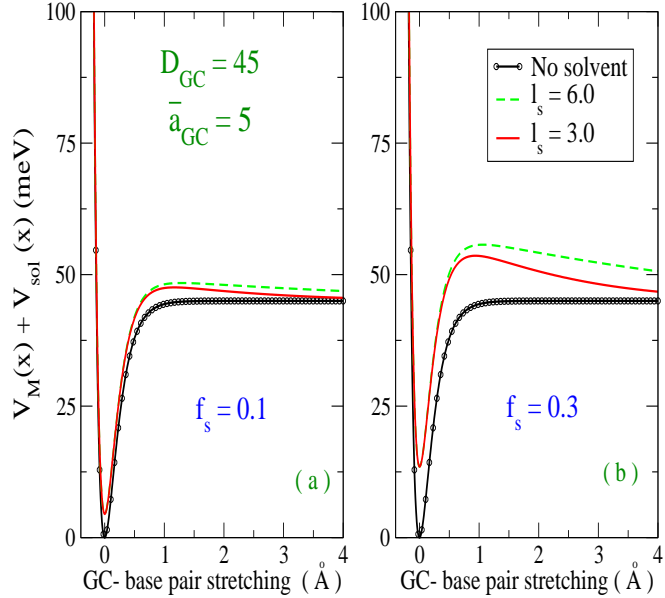}
\caption{\label{fig:4}(Color online)  Sum of Morse $V_M$ and solvent $V_{sol}$ potentials versus \emph{guanine-cytosine} base pair separation for two values, in (a) and (b) respectively, of the solvent barrier factor $f_s$.  $l_s$ (in {\AA}) tunes the width of the solvent barrier. Also the bare Morse potential with its peculiar plateau is plotted. $D_{GC}$ is in units meV and $\bar{a}_{GC}$ is in \AA$^{-1}$ .
}
\end{figure}

\textit{(ii)}  DNA molecules are usually surrounded by water hence their physical properties depend on the salt concentration in the solvent \cite{owcz04}. Looking at the Morse potential $V_{M}[r_n]$, it appears that  for bp fluctuations much larger than the bare helix diameter i.e., $|r_n| \gg  R_0$, the pair mates would sample the flat part of the Morse potential (see Fig.~\ref{fig:2}) and, in principle, they could go far apart with no further energy cost. This situation however does not account for those recombination
events which instead may take place in solution and calls for corrections to the physical picture provided by $V_{M}$ as discussed at length in the Introduction. 
Furthermore, when a base gets out of the stack, it may form a hydrogen bond with the solvent and, in order to re-close, the base encounters an entropic barrier, not described by $V_M$.
Following these arguments, I have introduced in Eq.~(\ref{eq:29}) a one-particle solvent potential, $V_{Sol}[r_n]$ depending on the input parameters $f_s$ and $l_s$ \cite{druk}. This potential enhances by $f_s D_n$ (with respect to the Morse plateau) the height of the energy barrier above which the bp dissociates and introduces a hump whose width is tuned by $l_s$. This length defines the range within which $V_{Sol}$ is superimposed to the plateau of the Morse potential. 
These features are visualized in Fig.~\ref{fig:4} which plots the one particle potential $V_{M} + V_{Sol}$ as a function of the bp distance, assuming $R_0=\,0$ and Morse parameters suitable to a $GC$ bp.
A broader discussion of the solvent effects on the DNA thermal properties may be found e.g. in ref.\cite{io11b}.

\textit{(iii)} Since all potential parameters are taken as site dependent, Eq.~(\ref{eq:29}) can be applied to model also heterogeneous double stranded sequences. The stacking parameters clearly refer to the two bps forming the dimer.

In order to extract predictions from the  Hamiltonian model for the structural and thermodynamic properties of the molecules, we have now to build a 
computational method. This is the subject of the next Section.

\section*{4. Computational Method}

\renewcommand{\theequation}{4.\arabic{equation}}
\setcounter{equation}{0}

As seen above, the distance between the complementary mates of any bp in the chain constantly fluctuates due to the dynamical interactions with the surrounding environment. Following this observation, I have developed a method based on the idea that the bp distances $r_n$ in Fig.~\ref{fig:3} can be conceived as time dependent trajectories and, accordingly, can be mapped onto the time scale: $r_n \rightarrow |r_n(\tau)|$ with $\tau$ being the imaginary time defined by $\tau=\,it_r$ and $t_r$ is the real time for the evolution amplitude of the particle trajectory within the time interval, $t_r(b) - t_r(a)$. 

The theoretical grounds of the method lie in the analytic continuation of the quantum mechanical partition function to the imaginary time axis which, in general, permits to obtain the quantum statistical partition function \cite{fehi}. Accordingly $\tau$ varies in a range $\tau(b) - \tau(a)$ whose amplitude is set by the inverse temperature $\beta$ and the partition function is written as an integral over closed trajectories, $\,r_n(0)=\, r_n(\beta) \,$,  running along the $\tau$-axis. 

The imaginary time formalism is widely used in semi-classical methods for the solution of quantum statistical problems  and it has been applied over the years to a number of condensed matter physics models, see for instance refs. \cite{io97,io04}. 

More recently the method has been adapted  to treat the ensemble of DNA molecules (which is a classical system usually considered at room temperature) as extensively described in refs.\cite{io14c,io14b,io16b}. While the reader may find in refs.\cite{fehi} the fundamentals of the path integral formalism,  the main features and equations useful to our purposes are hereafter outlined.

As a consequence of the above mentioned $\tau$-closure condition,  the $r_n(\tau)$ can be written in Fourier series around the average helix diameter:

\begin{eqnarray}
& &r_n(\tau)=\, R_0 + \sum_{m=1}^{\infty}\Bigl[(a_m)_n \cos(\omega_m \tau ) + (b_m)_n \sin(\omega_m \tau ) \Bigr] \, , \nonumber
\\
& &\omega_m =\, \frac{2 m \pi}{\beta} \, \, ,
\label{eq:30}
\end{eqnarray}

whereby a set of Fourier coefficients, $\{(a_m)_n,  (b_m)_n\}$, corresponds to a state for the n-th bp and provides a measure of the fluctuational distance between the complementary mates\footnotemark{}. 

\footnotetext{The coefficients $a_m$ should not be confused with the inverse length $\bar{a}_n$ of the site dependent Morse potential.} 

The expansion in Eq.~(\ref{eq:30}) defines the associated integration measure $\oint {D}r_n$  over the space of the Fourier coefficients:

\begin{eqnarray}
& &\oint {D}r_{n} \equiv  \prod_{m=1}^{\infty}\Bigl( \frac{m \pi}{\lambda_{cl}} \Bigr)^2 \int_{-\Lambda_{n}(T)}^{\Lambda_{n}(T)} d(a_m)_n \int_{-\Lambda_{n}(T)}^{\Lambda_{n}(T)} d(b_m)_n \, \, , \, \nonumber
\\
\label{eq:31}
\end{eqnarray}

where $\lambda_{cl}$ is the classical thermal wavelength (see ref.\cite{io14c})  and $\Lambda_{n}(T)$ is the temperature dependent cutoff for the radial fluctuations of the $n-th$ bp. 
The latter cutoff can be consistently determined by exploiting the normalization property intrinsic to the path integration technique \cite{io11a},  i.e. the condition that the measure in Eq.~(\ref{eq:31}) normalizes the kinetic term in the action :

\begin{eqnarray}
\oint {D}r_n \exp\Bigl[- \int_0^\beta d\tau {\mu \over 2}\dot{r}_n(\tau)^2  \Bigr] = \,1 \, \, .
\label{eq:31a} \,
\end{eqnarray}

Using  Eqs.~(\ref{eq:30}),~(\ref{eq:31}), the l.h.s. of Eq.~(\ref{eq:31a}) transforms into a product of independent Gau{\ss}ian integrals which can be solved by setting \, $\Lambda_{n}(T)=\,{{U_{n} \lambda_{cl}} / {m \pi^{3/2}}}$, with  $U_{n}$ being a dimensionless parameter. It is numerically found that Eq.~(\ref{eq:31a}) is satisfied by chosing $U_{n}=\,2$.

Importantly, it is also noticed that Eq.~(\ref{eq:31a}) holds for any $\mu$. This amounts to say that the system free energy does not depend on $\mu$, as expected for a classical system.
Moreover, the measure in Eq.~(\ref{eq:31}) permits to integrate both kinetic and potential actions over the same degrees of freedom thus avoiding the decoupling between momenta and real space integrations operated in the usual approach to the classical partition function, see Eq.~(\ref{eq:0b}) and the ensuing discussion. Then, Eq.~(\ref{eq:31}) correctly renders a dimensionless total partition function without the need to add an \textit{ad hoc} normalization constant as done in Eq.~(\ref{eq:0b}).

It follows that the general partition function associated to Eq.~(\ref{eq:29}) is given by:

\begin{eqnarray}
& &Z_N=\, \oint Dr_{1} \exp \bigl[- A_a[r_1] \bigr]   \prod_{n=2}^{N}  \int_{- \phi_{max} }^{\phi_{max} } d \phi_n \int_{- \theta_{max} }^{\theta _{max} } d \theta_{n} \oint Dr_{n}  \exp \bigl[- A_b [r_n, r_{n-1}, \phi_n, \theta_n] \bigr] \, , \nonumber
\\
& &A_a[r_1]= \,  \int_{0}^{\beta} d\tau H_a[r_1(\tau)] \, , \nonumber
\\
& &A_b[r_n, r_{n-1}, \phi_n, \theta_n]= \,  \int_{0}^{\beta} d\tau H_b[r_n(\tau), r_{n-1}(\tau), \phi_n, \theta_n] \, \, ,
\label{eq:32}
\end{eqnarray}

where $\phi_{max}$ and $\theta _{max}$ are the maximum amplitudes for the bending and twisting fluctuations which can be set in accordance with the experimental indications for specific molecules. 

It is remarked that:

\textit{i)} by virtue of the imaginary time mapping, the inverse temperature is introduced in the formalism. Accordingly, the bp fluctuations do depend on the temperature as it is expected on general grounds. 

\textit{ii)} As mentioned in Section 2.1,  the partition function of the PB (and PBD) model is customarily computed in Transfer Integral methods by applying periodic boundary conditions which amount to close the linear chain into a loop \cite{kalos20}. This procedure is however questionable in short chains due to the relevance of finite size effects. This drawback is avoided in the path integral formalism. In fact, the closure condition for the bp fluctuations is imposed here on the time axis whereas the chain maintains the open ends in real space.  Hence, there is no need to impose fictitious periodic boundary conditions.

\textit{iii)} The free energy of the system is computed  as: $F=\, -\beta ^{-1} \ln Z_N$. Then, the thermodynamical properties of a helical molecule in a solvent are derived from Eqs.~(\ref{eq:29}),~(\ref{eq:30}),~(\ref{eq:31}),~(\ref{eq:32}).

\subsection*{4.1 \, Radial Cutoff: Theory }

As shown above,  the maximum amplitude for the bp fluctuations can be technically determined, in the path integral method, by the normalization condition  for the free particle action. This mathematical condition clearly holds both for the PBD ladder model and for the DNA helical model
in Eq.~(\ref{eq:29}) and yields the minimal $U_{n}$ such that Eq.~(\ref{eq:31a}) is fulfilled. However, in order to compute specific properties of nucleic acids one may need to take a cutoff larger than the value set by the normalization condition.
In fact, larger cutoffs may be required to include those large amplitude fluctuations which affect the flexibility of the chain.  This points to the importance of defining a rigorous physical criterion which restricts the bp configuration space selecting a cutoff consistently with the model potential. 

In general, it can be reasonably assumed that the radial cutoff may also vary with the specific helical conformation of the molecule although, for the task of establishing a consistent relation between model parameters and radial cutoff, the details of the bp fluctuations over the twist and bending angles make a minor contribution. 

{Accordingly the angles $\phi_n$ and $\theta_n$ in Eq.~(\ref{eq:29}) are replaced by average values $\bar \phi$ and $\bar \theta$ which are taken as input parameters. This permits to retain the 3D nature of the model while the computational time for the following calculations is markedly reduced.}
Further, by tuning $\bar \phi$ and $\bar \theta$, one can study the relation between radial cutoff and macroscopic helical conformation. 

The latter is generally characterized by a given number of particles per helix turn. In the case of dsDNA (or dsRNA), let's say $h$ the number of bps per helix turn. This number is also named in the literature as the \textit{helical repeat}.   While $h$ may depend on temperature, salt concentration in the solvent, sequence length and specificities, the value usually reported for kilo-base long DNA  under physiological conditions is $h \sim \, 10.5$ \cite{depew}. This has to be understood as an average value considered that conformational fluctuations and buffeting of the solvent bath may locally distort the helix and change $h$.

Hereafter the helical repeat of the molecule is defined by, \, $h=\, 2\pi / \bar \theta$ .

To pursue our task, it is noticed that the bps of DNA in solution are constantly subjected to thermal fluctuations
which deform the molecular bonds causing transient openings along the chain. Accordingly, the bp thermal fluctuations are an example of constrained Brownian motion for a particle subjected to the specific interactions which stabilize the double helix.

Let's focus on the mid-chain $j-$th bp, for instance the A blue dot in Fig.~\ref{fig:3} and assume that, at the initial time, the average distance between the pair mates is, $< r_j > =\,R_0$. At any successive time $t$, fluctuations may cause $r_j$ to contract or expand with respect to $R_0$. Accordingly,  $P_j(R_0,\, t)$ is  defined as the probability  that $r_j$ does not return to the initial value up to $t$ and $F_j(R_0,\, t)=\, - d P_j(R_0,\, t) dt$ as the probability that the {path will return to the origin for the first time } between $t$ and $t + dt$.

For the $j-th$ bp embedded in the chain and interacting with its first neighbors via the stacking potential in Eq.~(\ref{eq:29}), I write $P_j(R_0,\, t)$ as a path integral :

\begin{eqnarray}
& &P_j(R_0,\, t)=\,   \oint Dr_{1} \exp \bigl[- A_a[r_1] \bigr] 
\cdot \prod_{n=2, \, n\neq j}^{N} \oint Dr_{n}  \exp \bigl[- A_b [r_n, r_{n-1}, \bar \phi, \bar \theta] \bigr]  \cdot \,  \nonumber 
\\
& & 
\int_{r_j(0)}^{r_j(t)} Dr_{j} \exp \bigl[- A_b[r_j, r_{j-1}, \bar \phi, \bar \theta] \bigr]  \cdot
\prod_{\tau=\,0}^{t}\Theta\bigl[r_j(\tau) - R_0\bigr] \, \, , \nonumber 
\\
\label{eq:33}
\end{eqnarray}

where the Heaviside function $\Theta[..]$ enforces the condition that $r_j(\tau)$ has to remain larger than $R_0$ for any $\tau \in [0, t]$.
This is implemented in the code by evaluating at any imaginary time \, $\tau$ the amplitude of $r_j(\tau)$ in Eq.~(\ref{eq:30}) and discarding those sets of coefficients which don't comply with such condition.
The need to introduce two time variables, $t$ and $\tau$, arises from the fact that, for a given $t$, the probabilities are given as a sum over the particle histories $r_j(\tau)$ in the time lapse  $[0, t]$ \cite{maj05}. Also note that $t$ here is the upper bound for $\tau$ along the imaginary axis and should not be confused with $t_r$ defined at the beginning of Section 4. 
 
Moreover, the measures of integrations over closed ($\oint {D}r_{n}$) and open  ($\int Dr_{j}$) trajectories in Eq.~(\ref{eq:33}) are coupled via the two particle potential which connects the $n-th$ and $j-th$ bps along the stack.
The actions  in Eq.~(\ref{eq:33}) are obtained by the following $d\tau$ integrals:

\begin{eqnarray}
& &A_a[r_1]=\,\int_{0}^{\beta} d\tau H_a[r_1(\tau)] \, , \nonumber 
\\
& &A_b [r_n, r_{n-1}, \bar \phi, \bar \theta]=\, \int_{0}^{\beta} d\tau H_b[r_n(\tau), r_{n-1}(\tau), \bar \phi, \bar \theta] \, , \nonumber 
\\
& &A_b [r_j, r_{j-1}, \bar \phi, \bar \theta]=\, \int_{0}^{t} d\tau H_b[r_j(\tau), r_{j-1}(\tau), \bar \phi, \bar \theta] \, ,
\label{eq:34}
\end{eqnarray}

whereby it is pointed out that the functional for the $n-$ bps is an integral over closed trajectories (as in Eq.~(\ref{eq:32})) whereas the functional for the $j-$ bp is an integral over open trajectories.

As a consequence, for the $j-th$ bp, a Fourier series expansion as in Eq.~(\ref{eq:30}) can be still performed, but the normalization condition in Eq.~(\ref{eq:31a}) cannot be applied. This follows from the observation that, for any $\tau$,   $r_j(\tau)$ is defined up to $r_j(t)$ which is in fact an open trajectory for any $t < \beta$. Hence, a new criterion should be developed to estimate the integral cutoff on the amplitude of the $j-th$ radial fluctuation.

The criterion is built by inspecting Eq.~(\ref{eq:33}) and asking the question: what is the probability that, at the initial time, the $j-th$ fluctuation is larger than $R_0$ ?  

From Eq.~(\ref{eq:30}), at $t=\,0$, the $j-th$ trajectory is \, $r_j(0)=\, R_0 + \sum_{m=1}^{\infty}(a_m)_j$ with the Fourier coefficients being integrated on an even domain. Accordingly, the initial probability $P_j(R_0,\, 0)$ is expected to be $ \sim 1/2$ \footnotemark{}. 

This is the constraint which needs to be fulfilled in the computation of the first-passage probability as a function of time. In order to implement this criterion, the Fourier integration \, $\int_{r_j(0)}^{r_j(t)} Dr_{j}$  in Eq.~(\ref{eq:33}) is integrated by setting a cutoff \, $\Lambda_{j}(T)=\,{{U_j \lambda_{cl}} / {m \pi^{3/2}}}$, with tunable $U_j$: then,  the precise value \, $U_j$ such that $P_j(R_0,\, 0) \sim 1/2$ is eventually selected.  In this way one picks the cutoff on the base of a robust physical constraint for a specific set of model parameters and for a given helical conformations. While the mid-chain bp has been chosen for reference, the method is general and holds for any bp in the chain.

\footnotetext{In principle the Fourier coefficients in Eq.~(\ref{eq:31}) are integrated on an even domain. However too negative $a_m$'s are discarded due to the physical condition associated to the hard core of the one particle potential. The latter is tuned by the parameter which regulates the range of the Morse potential (see Section 2). The asymmetry in the choice of $a_m$'s included in the computation explains why $P_j(R_0,\, 0)$ may get slightly larger than $1/2$. Hence the approximation sign used in the text.
}

\subsection*{4.2 \, Radial Cutoff: Results }

The theory is now tested for some specific cases by setting in Eq.~(\ref{eq:33}) the average bending at $\bar \phi=\,6^{o}$ consistently with the indications of Fluorescence Resonance Energy Transfer studies probing the DNA bending elasticity  at short length scales \cite{kim14}. 
The average twist angle is initially taken as $\bar \theta=\,36^{o}$, which corresponds to an average helical repeat \, $h=\,10$,  close to the usual experimental value for kilo-base long DNA chains at room temperature. Although the model potential contains only first neighbors radial interactions, the bps are correlated along the stack due to the helical conformation. Here I am taking a short homogeneous chain of $N=\,21$ bps which allows for about two turns of the helix. The diameter is $R_0=\, 20$ \AA. The bare rise distance $d$ is, at first, set equal to zero like in the PBD ladder model.

The potential parameters are those taken in ref.\cite{io09} namely,  $D_{n}=\,30$ meV, $a_{n}=\,4.2$ \AA$^{-1}$, $K_{i}\equiv K_{n,n-1}=\,60$ meV $\cdot$ \AA$^{-2}$, 
$\rho_{n}\equiv \rho_{n,n-1} =\,1$,  $\alpha_{n}\equiv \alpha_{n,n-1} =\,0.35$ \AA$^{-1}$. This set is consistent with the parameters  used by other groups  in investigations of the PBD model \cite{zhang97,campa98} and derived by fitting the experimental melting temperatures, though some discrepancies persist mostly as for the stacking force constants \cite{eijck11,io21}.  It is also pointed out that the terminal bps of a chain lack a first neighbor and generally tend to unstack. To simulate this effect which is all the more relevant in short chains, the stacking parameters \, $K_{2,1}$ and $K_{N,N-1}$ are taken one half of the value assumed for the internal dimers. While this choice is arbitrary, it permits to weigh the impact of chain end effects on the bp cutoff.

Then, we are ready to calculate the probability in Eq.~(\ref{eq:33}) as a function of time for different $U_j$ and select the good cutoff, \, $U_j \equiv \bar{U}$, such that \, $P_j(R_0,\, 0) \sim 1/2$ for all internal bps (in view of the fact that the chain is homogeneous).  
Instead, for the terminal bps, the first passage probability remains smaller than $1/2$ as the average bp separation remains larger than $R_0$. This result follows from the assumption of softer stacking force constant for the terminal dimers. This causes looser bonds and fraying at the chain ends.

\begin{figure}
\includegraphics[height=7.0cm,width=8.0cm,angle=-90]{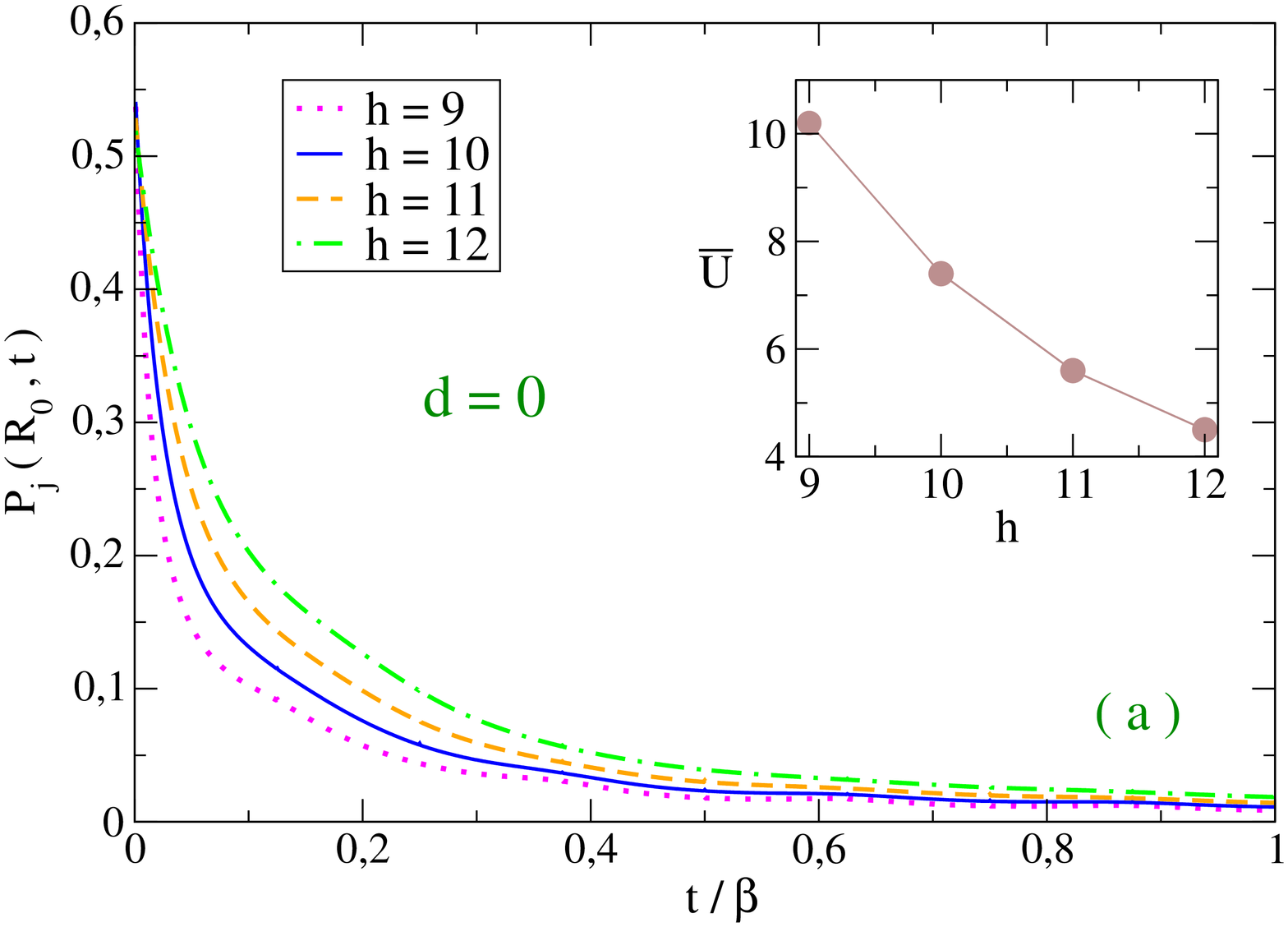}
\includegraphics[height=7.0cm,width=8.0cm,angle=-90]{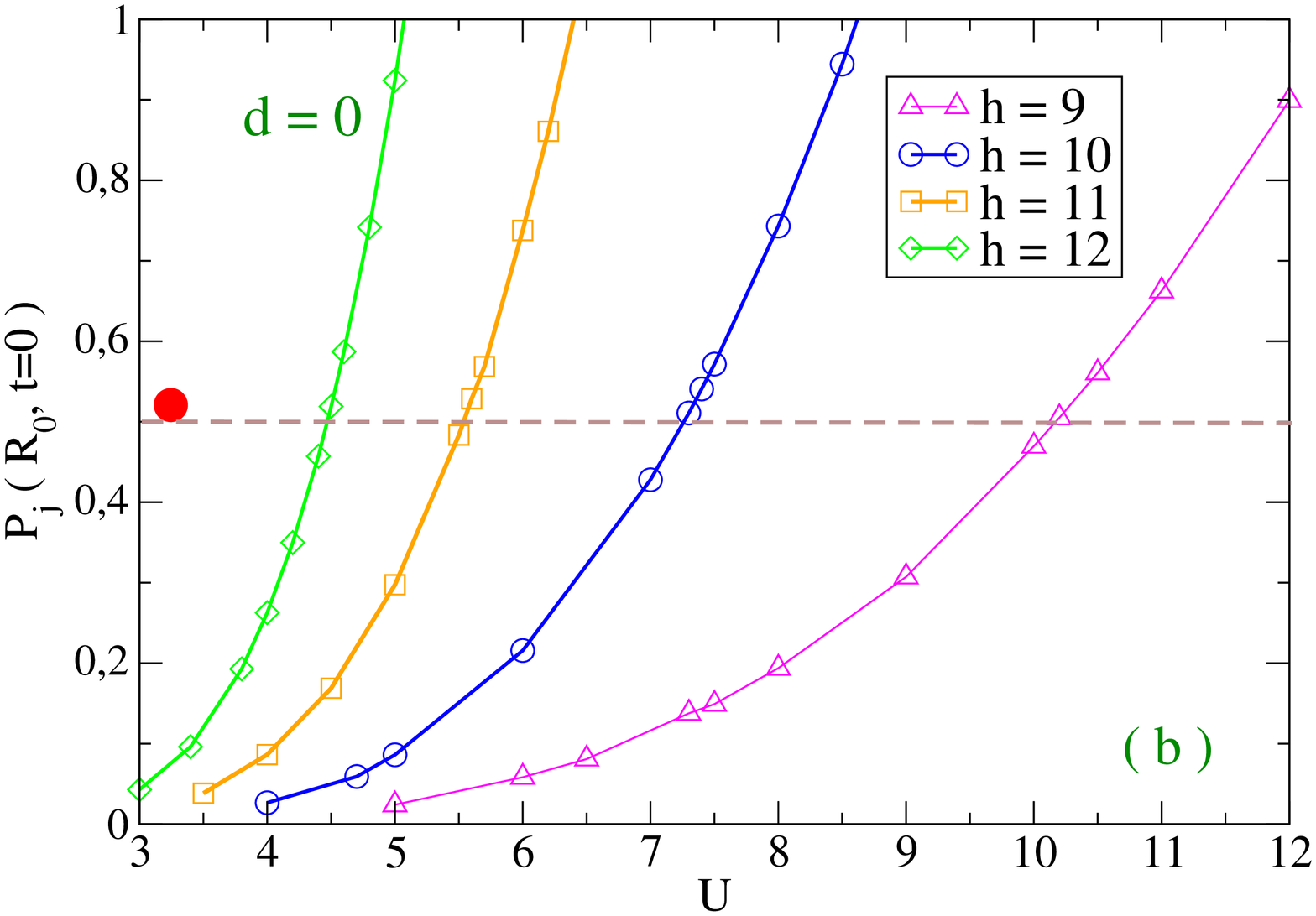}
\caption{\label{fig:5}(Color online)  
(a) First-passage probability versus time for the mid-chain bp in equilibrium with the $N-1$ bps at room temperature.  A homogeneous chain is taken with $N=\,21$. Four $h$ values are considered. {The respective average twist angles $\bar \theta$  are (from top to  bottom in the legend) \, $40^o, \,36^o, \, 32.7^o, \, 30^o $.}
 For each twist conformation, the probability is computed assuming the respective cutoff \={U}
(in the inset) that fulfills the initial time condition. (b) Zero time probabilities versus integral cutoff for three twist conformations. The $U$ values for which the plots intersect the dashed line, are the $\bar U$'s reported in the inset in (a).  The red dot {\color{red} $\large \bullet$ } \,  marks the cutoff found for the PBD ladder model. }
\end{figure}

The probability is evaluated over 1000 points along the time axis. Hence the \textit{zero time} value $P_j(R_0,\, 0)$ corresponds to the abscissa $t / \beta =\,10^{-3}$.

For practical purposes, only the first Fourier component ($m=\,1$) needs to be taken for the path expansion in Eq.~(\ref{eq:30}). This suffices to achieve numerical convergence in the computation of Eq.~(\ref{eq:33}) and, generally, also of Eq.~(\ref{eq:32}). 

The results are displayed in Fig.~\ref{fig:5}. The $P_j(R_0,\, t)$'s in 
Fig.~\ref{fig:5}(a) are calculated by taking the $\bar{U}$'s obtained respectively from the plots in Fig.~\ref{fig:5}(b). As a main result, $\bar{U}$ markedly decreases for larger $h$. This is understood by observing that our helical chains are considered to be stable at room temperature also in the untwisted conformations although, in the latter, large amplitude bp fluctuations would easily disrupt the hydrogen bonds and unstack the helix. Accordingly, helical molecules in a large $h$ conformation sustain only short scale fluctuations in order to preserve the overall stability. This interpretation however ensues from the simplyfing assumption of a model with no intrinsic stiffness i.e., with $d=\,0$.  By further increasing $h$ the helix unwinds and tends to the ladder representation. 
Consistently, it is shown in Fig.~\ref{fig:5}(b), that the selected $\bar{U}$'s tend to the cutoff value determined for the PBD model. As we have seen in Section 2, the PBD model in fact takes $d=\,0$ .

\begin{figure}
\includegraphics[height=7.0cm,width=8.0cm,angle=-90]{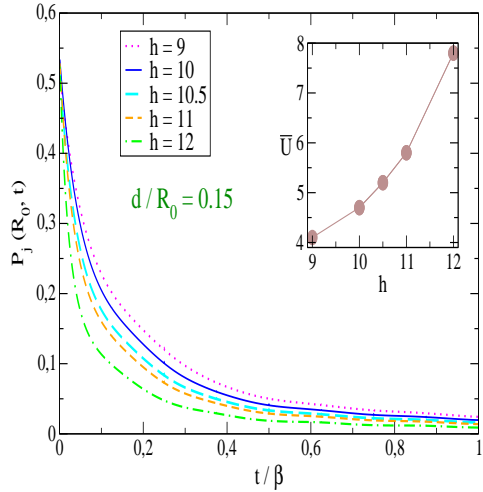}
\caption{\label{fig:6}(Color online)  
As in Fig.~\ref{fig:5}(a) but with a finite rise distance $d$.  Five twist conformations are considered. {The respective average twist angles $\bar \theta$  are (from top to  bottom in the legend) \, $40^o, \,36^o, \, 34.3^o, \, 32.7^o, \, 30^o $.} For each $h$ value, the first passage probability is computed assuming the respective cutoff \={U} (in the inset) that fulfills the initial time condition.  }
\end{figure}

The picture substantially changes once  a finite rise distance $d$ is introduced  to yield a more realistic model for the helical molecule. The results are shown in Fig.~\ref{fig:6} where the time dependent probability is computed as a function of time by varying the average twist angle. Now the intrinsic stiffness  $d$ confers stability to the helix which can sustain large amplitude bp fluctuations also in the untwisted conformations. Accordingly the 
cutoff value, for which the zero time probability condition is fulfilled, grows versus $h$ as shown in the inset. For instance, given a chain with $h=\,10.5$, the obtained value $\bar{U} =\,5.2$ yields a maximum amplitude $\Lambda_{j}(T)=\,1.08$ \AA \, for the first Fourier component in Eq.~(\ref{eq:30}). This, in turn, corresponds to a reasonable estimate of $\sim 2.2$ \AA \, for the largest breathing fluctuation of the $j-th$ bp with respect to the average helix diameter in the closed state. This length is $\sim 10 \%$ of $R_0$ and it is a fair measure for the threshold above which hydrogen bonds are disrupted.

Although the discussion carried out so far has focused exclusively on DNA molecules, I wish to emphasize that both the model and the method can be likewise applied to the sister molecule, the double stranded RNA, after introducing some modifications which account for its structural peculiarities. 

With this caveat and having established a method to determine the radial cutoff, we are now in the position to calculate some physical properties of molecules in solution.

\section*{5. Applications }

The computational techniques presented in Section 4 have been applied over the years to evaluate several properties both of homogeneous and heterogeneous 3D DNA chains such as cyclization, persistence lengths, distribution lengths, mechanical stretching and end-to-end distance (see  Fig.~\ref{fig:5}) both for free molecules and in confining environments \cite{io17,io18,io18b,io18c,io16,io19,io20b}.
 
As an example let's focus here on the cyclization probability, also named the $J$-factor,  that is the probability for the occurrence of the circular conformation  given an ensemble of molecules in the linear form. While this property is a strong indicator of the chain flexibility, there has been a renewed interest towards it following some experiments \cite{vafa,kim13} which have found a $J$-factor much larger in short chains than that predicted by the standard worm-like-chain model (WLC) \cite{stock72}. The $J$-factor can be computed by Eq.~(\ref{eq:32}) after building the fraction of molecules whose terminal bps happen to be within a capture volume, that is very close to each other. The details of the calculation performed for short chains are in ref.\cite{io16b}. The $J$- factor turns out to be strongly dependent on the potential parameters and specifically on the force constants $\rho _n$ and $\alpha _n$.  Accordingly
one may fit the model predictions to the experimentally available cyclization data in order to determine consistently the non-linear stacking parameters. 

\begin{figure}
\includegraphics[height=8.0cm,width=8.0cm,angle=-90]{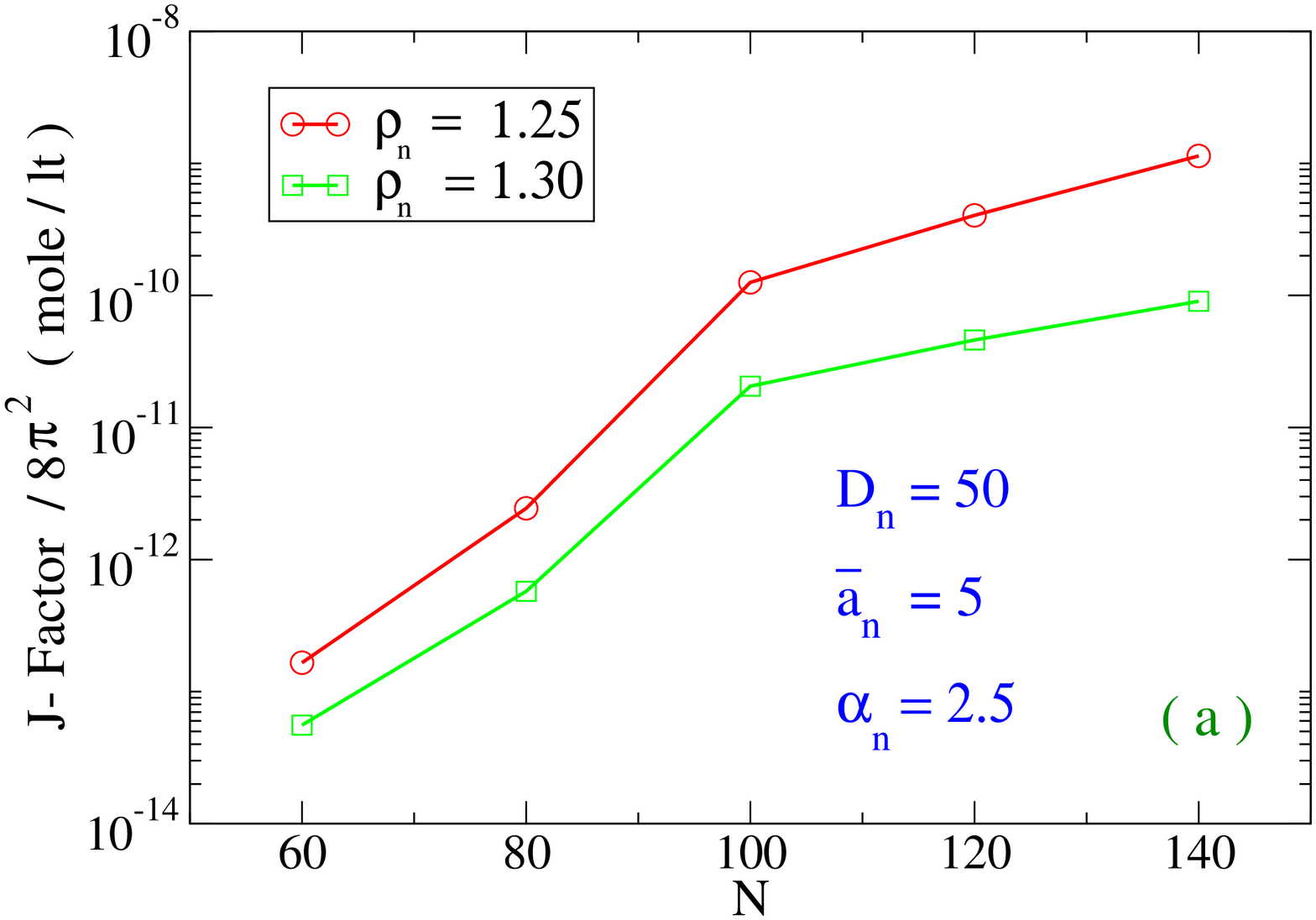}
\includegraphics[height=8.0cm,width=8.0cm,angle=-90]{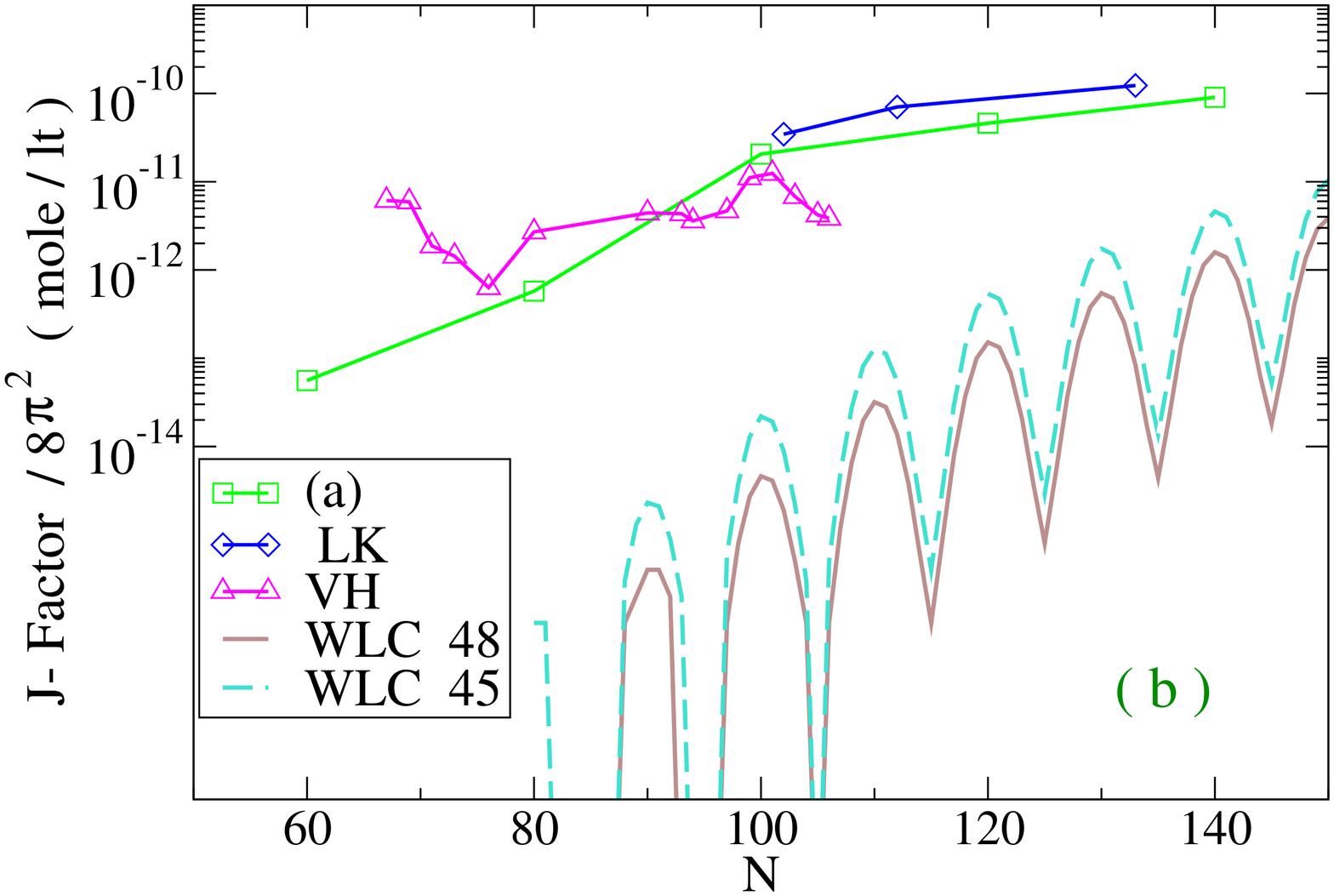}
\caption{\label{fig:7}(Color online)  
(a) $J$- factor (over $8 \pi^2$) calculated for a set of five sequence lengths ($N$).   The potential parameters refer to homogeneous sequences and are the same for all $N$'s. Two values of the non-linear stacking stiffness $\rho _n$ are taken. (b) The green plot in (a) is compared to the experimental results of ref.\cite{vafa} (VH) and ref.\cite{kim13} (LK). The $J$-factor of the twisted worm-like-chain model (WLC) \cite{shimada}  is computed for two persistence lengths, $45$ nm and $48$ nm.   }
\end{figure}

In particular, I have considered the cyclization of single DNA molecules yielding a $J$- factor  $\sim  10^{-9}$ mol / liter for $N \sim 100$  as measured by FRET. For this length scale, independent experiments report consistent values  as shown in Fig.~\ref{fig:7}(b) . 

Then, I have set a pair of parameters, e.g. $\alpha _n=\,2.5$ \AA$^{-1}$ and $\rho _n=\,1.3$, which fit the mentioned $J$- factor value and computed the $J$- factor as a function of the molecule length, for five $N$ values, as shown in Fig.~\ref{fig:7}(a). 
Also the case with $\rho _n=\,1.25$ is considered to remark the strong dependence of the cyclization probability on the molecule stiffness.

In this calculation the helical repeat has been taken constant, $h=\,10$. It follows that  $N / h$ is always an integer for the considered chain lengths and no extra twist is necessary to close the chain into a loop. Under this conditions, the molecule looping occurs at fixed helical repeat and does not require the unwinding of the complementary strands. 
For this reason, the peculiar oscillations of the $J$- factor, due to the twist rigidity of the double helix, do not appear in the plots. 

The $J$- factor drops by decreasing $N$, markedly below $N=\,100$, in accordance with the qualitative general expectations.  
However such drop is not so abrupt as predicted by the WLC model reported in Fig.~\ref{fig:7}(b) for two values of persistence length \cite{shimada}. In fact, the computed $J$- factor is still sizeable i.e. $\sim 10^{-11}$ at $N=\,80$.  While the plots in Fig.~\ref{fig:7}(a) refer to homogeneous chains, the displayed trend (sequence length dependence) and the body of our results would not be modified by heterogeneity effects. Not even different choices for the pair ($\rho _n$, $\alpha _n$) would change such trend. 

The results are compared in Fig.~\ref{fig:7}(b) to the mentioned FRET experiments made at very short length scales \cite{vafa,kim13}. Indeed, some relevant differences exist between the two sets of data indicating the difficulty in extracting $J$- factors from experiments  and in performing quantitative comparison between models and data.
Nevertheless both sets of FRET data concur that there is a sizeable cyclization probability at very short molecule lengths and, importantly, our theoretical model can predict this feature for a consistent choice of the potential parameters although the oscillations in the $J$-factor experimental plot are not reproduced for the reasons explained above. 

I have chosen to focus on the looping probability as this property provides a relevant benchmark for the theory and can be suitably described only by models which account for the flexibility of the double helix at short length scales.
The structure of the stacking potential in Eq.~(\ref{eq:29}) which allows for large local bending fluctuations and the specific integration technique which includes a broad ensemble of independent path fluctuations, both contribute to shape a model for the helix with flexible hinges at the level of the bp. These mechanisms are responsible for the substantial molecule bendability which leads to the results shown in Fig.~\ref{fig:7}.

\section*{6. Conclusions }

Nucleic acids are macromolecules containing a huge number of atoms even for short sequences made only of a few tens of base pairs. In dealing with these double helical molecules, theorists often turn to mesoscopic models which describe the fundamental intra-strand and inter-strand forces at the level of the base pair and permit to get meaningful predictions for the thermodynamical and structural properties.
Beginning with a well-known one-dimensional and non-linear model, proposed long ago to investigate the DNA melting, I have shown how the level of complexity of the Hamiltonian can be increased by progressively adding new ingredients i.e., new degrees of freedom, thus providing a more accurate description of the double helix in three dimensions. 
While the 3D model presented in Section 3 importantly accounts for the twisting and bending fluctuations between neighboring base pairs along the molecule axis, still it provides a coarse-grained picture for the molecule: in fact it assumes a point-like description for the nucleotide made of a nitrogenous base, a sugar ring and a phosphate group. Further, it neglects the distortion of the hydrogen bonds between complementary bases.  Nevertheless our 3D Hamiltonian description contains sufficient structural details to investigate the flexibility of helical chains at short length scales. Certainly, it should be always considered that any step upward in the hierarchy of model complexity also entails an increase of the computational time required to extract physical information from that specific model.
Finally I have presented the main properties of a powerful computational technique, based on the path integral formalism, which has been widely used to derive various physical properties of short sequences both in the linear and circular form. I have also discussed in detail the statistical method applied to determine the cutoff on the amplitude of the base pair fluctuations. While the applications of the theory have generally regarded DNA chains, some latest work has shown possible pathways to extend the analysis to double stranded RNA molecules.


\end{document}